\documentclass[
reprint,
aps,
superscriptaddress,
nofootinbib,
nobibnotes,
twocolumn]{revtex4-2}

\usepackage{xcolor}
\usepackage{graphicx}
\graphicspath{{./fig/}}
\usepackage{amsmath,amssymb,amsfonts,mathrsfs,bm}
\usepackage{slashed}
\usepackage[%
colorlinks=true,
linkcolor=blue,
citecolor=blue,
]{hyperref}
\usepackage{braket}
\usepackage{makecell}  
\usepackage{subfigure}  

\begin{document}
\title{Probing Warm Inflation via Correlated Gravitational Waves from  First Order Phase Transitions}

\author{Xiao-Bin Sui}
\email{suixiaobin21@mails.ucas.ac.cn}
\affiliation{School of Fundamental Physics and Mathematical Sciences, Hangzhou Institute for Advanced Study, University of Chinese Academy of Sciences (HIAS-UCAS), Hangzhou 310024, China}
\affiliation{CAS Key Laboratory of Theoretical Physics, Institute of Theoretical Physics, Chinese Academy of Sciences, Beijing 100190, China}
\affiliation{University of Chinese Academy of Sciences, Beijing 100049, China}

\author{Jing Liu}
\email{liujing@ucas.ac.cn}
\affiliation{International Centre for Theoretical Physics Asia-Pacific, University of Chinese Academy of Sciences, Beijing 100190, China}
\affiliation{Taiji Laboratory for Gravitational Wave Universe (Beijing/Hangzhou), University of Chinese Academy of Sciences, Beijing 100049, China}

\author{Rong-Gen Cai}
\email{cairg@itp.ac.cn}
\affiliation{Institute of Fundamental Physics and Quantum Technology, Ningbo University, Ningbo, 315211, China}

\begin{abstract}
 

We investigate the properties of gravitational waves generated by heating-induced phase transitions in warm inflation. In this scenario, the heating phase of inflation followed by subsequent cosmological cooling can trigger two associated first-order phase transitions and generate characteristic gravitational waves. The correlated gravitational wave spectral features—amplitude, peak frequencies, and oscillatory behavior—originate from a unified model governing both phase transitions. These signatures allow discrimination between warm and cold inflation models, and give constraint on the key parameters including the dissipative coupling strength and the inflationary energy scale, collectively illuminating early-Universe dissipative dynamics. Future gravitational wave observatories such as BBO, Ultimate-DECIGO, $\mu$Ares, resonant cavities, and Pulsar Timing Array experiments, will play a important role in testing these theoretical
predictions.

\end{abstract}

\maketitle

\section{Introduction}
The inflationary paradigm has long been the cornerstone of modern cosmology, providing a compelling framework for explaining the large-scale homogeneity, flatness, and the origin of primordial density fluctuations in the Universe~\cite{Guth:1980zm,Guth:1980zm,Planck:2018jri,Mukhanov:1990me,Baumann:2009ds}. 
In the standard cold inflation (CI) scenario, a nearly isolated inflaton field drives the quasi-exponential expansion of the Universe, with particle production and reheating occurring only after inflation ends. 
However, this picture faces several theoretical challenges, including the requirement for an ultra-flat inflaton potential, $\eta$ problem~\cite{Copeland:1994vg,Arkani-Hamed:2003wrq}, and the exclusive reliance on quantum fluctuations as the source of primordial density perturbations. 
In contrast, warm inflation (WI)  introduces dissipative interactions between the inflaton and 
thermal bath during inflation~\cite{Berera:1995ie,Berera:2023liv,Kamali:2023lzq}, enabling concurrent particle production and radiation generation, and this mechanism alleviates the stringent constraints of CI, allowing the generation of classical thermal fluctuations, and a smooth transition to the radiation-dominated era without the need for a separate reheating phase~\cite{Berera:1995wh,Berera:1998px,Bartrum:2013fia,Berera:2008ar,Berera:2002sp,Berera:1998px,Motaharfar:2018zyb,Berera:2003yyp,Bastero-Gil:2016qru,Wang:2025duy}.

During inflation, the vast majority of the energy of the Universe is stored in the inflaton field, whereas the late Universe is characterized by a thermal bath that cools down due to the expansion of the Universe. 
A central challenge for any inflationary models is to explain the origin of this thermal bath. 
In CI, this issue is typically addressed by the reheating process, in which the inflaton decays into radiation after inflation ends~\cite{Kofman:1997yn,Kofman:1994rk,Bassett:2005xm,Shtanov:1994ce}. 
In WI, however, dissipative interactions between the inflaton and the thermal bath play a crucial role during the inflationary phase: the inflaton continuously transfers energy to the thermal bath, maintaining it at a nearly constant temperature throughout the inflationary epoch, rendering a distinct reheating phase unnecessary. 
This energy transfer can lead to transient increases in the temperature of thermal bath, potentially triggering a change in the vacuum state of a phase transition~(PT) field. 
The system then undergoes a heating phase transitions (hPTs), as it transitions to the true vacuum in response to the rising temperature, which have emerged as an important topic in cosmology recently~\cite{Buen-Abad:2023hex,Ai:2024cka,Barni:2024lkj}.
This contrasts with cooling phase transitions (cPTs), which typically occur in the post-inflationary, radiation-dominated era as the thermal bath cools.

When the hPT is first-order, it can generate a detectable stochastic gravitational-wave background (SGWB), carrying direct information about the nature of the early Universe from epochs otherwise inaccessible to electromagnetic telescopes.
The detection of SGWBs offer a unique window into the physics of the early Universe, as they can encode imprints of diverse high-energy phenomena such as reheating/preheating, phase transitions, topological defects, and primordial scalar perturbations~\cite{Cai:2017cbj,Bian:2021ini,Saito:2008jc,Ananda:2006af,Kohri:2018awv,Cai:2019amo,Cai:2018dig,Cai:2019cdl,Domenech:2021ztg,Lozanov:2019ylm,Liu:2023tmv,Bhaumik:2022pil,Sui:2024nip,Sui:2024grm,Zhou:2024yke,Inomata:2018epa,Liu:2023tmv,Fu:2023aab,Zeng:2023jut,Peng:2021zon,Li:2024lxx}.
During a first-order PT, the Universe undergoes a transformation from a metastable false vacuum state to a stable true vacuum state through the quantum nucleation of true vacuum bubbles and their subsequent rapid expansion~\cite{Coleman:1977py,Callan:1977pt,Linde:1980tt}. 
The complex dynamics of bubble collisions, sound waves, and magnetohydrodynamic turbulence in this process efficiently convert the latent energy into GWs~\cite{Jedamzik:1999am,Liu:2021svg,Liu:2022lvz,Ellis:2019oqb,Baker:2021nyl,Cai:2024nln,Flores:2024lng,Caprini:2015zlo,Caprini:2019egz}, making first-order PTs a key observational target for next-generation GW observatories such as LIGO/VIRGO/KAGRA~\cite{KAGRA:2021kbb}, Taiji~\cite{Ruan:2018tsw,Luo:2021qji}, TianQin~\cite{TianQin:2015yph}, Pulsar Timing Array~\cite{Xu:2023wog}, $\mu$Ares~\cite{Sesana:2019vho}, LISA~\cite{Barausse:2020rsu} and resonant cavities~\cite{Herman:2022fau}. 

In this paper, we investigate the generation of GWs during hPTs in the WI scenario. 
The exponential expansion of the Universe during inflation leads to the redshifting of the GW signal by a factor determined by the e-folds number between the hPT and the end of inflation. 
The energy spectrum of the generated GWs exhibits characteristic oscillatory features arising from the interplay between the thermal bath dynamics and the Hubble expansion, which are absent in the GWs produced by post-inflationary reheating in CI scenarios~\cite{An:2020fff,An:2022cce,Hu:2025xdt}. 
Furthermore, as the post-inflationary Universe cools, a symmetry-breaking first-order PT inevitably occurs from the same model, generating a second distinct peak in the SGWB which is expected to be detected by multiband GW observation methods.
These correlated distinctive signatures from this hPT-cPT sequence make the SGWB a powerful probe for distinguishing between CI and WI models, and constrain key parameters, including the strength of the dissipative coupling, the duration of inflation, and the inflationary energy scale. 

This paper is organized as follows. 
In Section \ref{sec2}, we presents the theoretical framework of WI, outlining the fundamental equations  and analyzing the temperature evolution of the heating process. 
Section \ref{sec3} proposes the PT model and analyzes the GW production with different parameters, highlighting the distinctions between hPTs and cPTs.
We conclude in Section~\ref{conclusion}. For convenience, we set $c=\hbar=1$ and the reduced Planck mass $M_{\text{pl}} = (8\pi G)^{-1/2} \approx 2.44 \times 10^{18} \,\text{GeV}$ throughout this paper.
\section{thermal history of warm inflation}
\label{sec2}
\subsection{The Model of Warm Inflation}
\label{subsec:IIA}
In WI, the inflaton field coexists with a thermal bath whose temperature \( T \) significantly exceeds the Hubble parameter \( H \) during inflation. Despite the high temperature, the energy density $\rho_{\text{r}}$ remains much smaller than that of the inflaton, i.e., $\rho_{\text{r}} \ll \rho_\phi$. To maintain the high temperature of the thermal bath during the quasi-exponential expansion, the inflaton must persistently transfer energy into the bath at a rate compensating for Hubble dilution, while efficient particle self-interactions within the bath are required to preserve thermal equilibrium.

The large coupling between the inflaton and  the radiation leads to significant dissipative effects, modifying inflationary dynamics beyond those single-field models. 
In WI, the evolution of the inflaton and radiation energy densities is governed by the following equations~\cite{Bastero-Gil:2009sdq}:


\begin{equation}
\label{fun1}
\ddot{\phi}+\left(3H+\Upsilon\right)\dot{\phi}+V_{,\phi}=0\,,
\end{equation}
\begin{equation}
\label{fun2}
\dot{\rho}_{\text{r}}+4H\rho_{\text{r}}=\Upsilon\dot{\phi}^2\,,
\end{equation}
where dots denote derivatives with respect to the cosmic time, $H = \dot{a}/a$ is the Hubble parameter, $V_{,\phi}\equiv\frac{d V}{d\phi}$,  and $\Upsilon$ is the dissipation coefficient arising from interactions between the inflaton and the thermal bath. 
For convenience, we define the dimensionless dissipative ratio $Q \equiv \Upsilon/(3H)$. The regime $Q > 1$ corresponds to ``strong dissipation'', whereas $Q < 1$ corresponds to ``weak dissipation''.

Under the slow-roll approximation, the Hubble slow-roll parameter $\epsilon_H$ can be expressed in terms of the potential slow-roll parameter $\epsilon_V$ as~\cite{Moss:2008yb}:
\begin{equation}
\epsilon_H \approx \frac{\epsilon_V}{1+Q}\,,
\end{equation}
where $\epsilon_V \equiv \frac{M_{\text{pl}}^2}{2} \left( \frac{V_{,\phi}}{V} \right)^2$. 
By combining Eq.~\eqref{fun1} and \eqref{fun2} and using slow-roll approximation, the ratio of radiation energy density to potential energy density is
\begin{equation}\label{eq:ratio}
\frac{\rho_{\text{r}}}{V} \approx \frac{1}{2} \epsilon_H \frac{Q}{1+Q}\,.
\end{equation}
The sustained duration of inflation requires $\epsilon_H \ll 1$, which ensures $\rho_{\text{r}} \ll V$ holds for both large and small values of $Q$. However, as inflation terminates (when $\epsilon_H \sim 1$), Eq.~\eqref{eq:ratio} indicates that $\rho_{\text{r}} \sim V$ in the strong dissipation regime. This implies that radiation naturally becomes a significant energy component by the end of inflation, facilitating a smooth transition to radiation domination without the need for an additional reheating phase, which is an advantage over the CI scenario. This thermalization mechanism is also well-realized in many WI scenarios, such as the axion-like warm inflation models~\cite{Das:2019acf,Laine:2021ego,DeRocco:2021rzv}.
Furthermore, we also note that the termination condition of WI depends on the specific form of the inflationary potential and the magnitude of the dissipation coefficient. 
Specifically, the persistence of inflation is guaranteed when \(\epsilon_V < 1 + Q\), while inflation terminates once \(\epsilon_V \approx 1 + Q\). 
Given that \(Q\) itself is a dynamical parameter, WI possesses multiple mechanisms for graceful exit.


\subsection{Temperature Evolution During the Heating Process}\label{sec:IIB}
\begin{figure*}
    \centering
    \begin{minipage}[t]{0.24\textwidth}
        \centering
        \includegraphics[width=\textwidth]{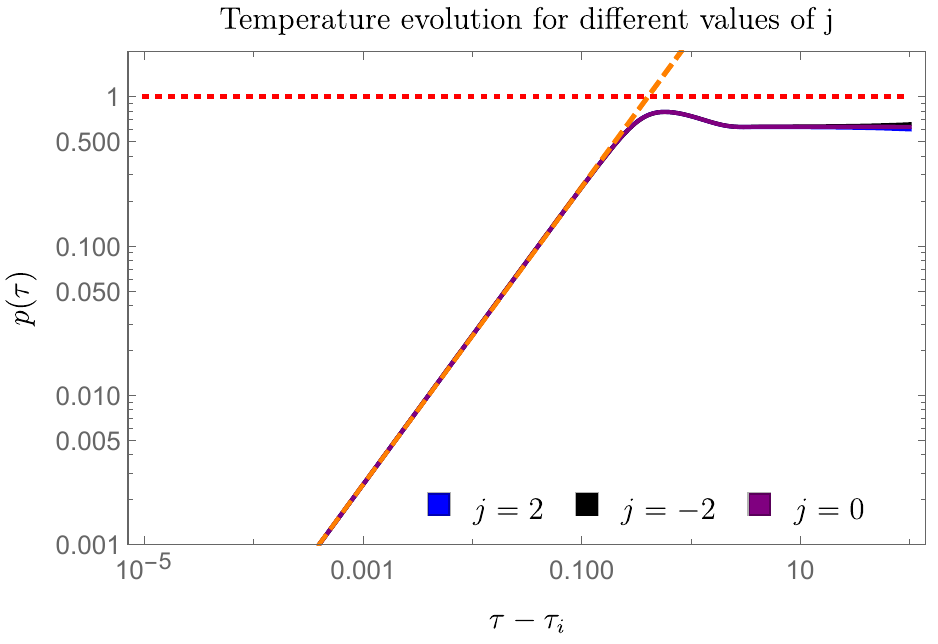}
        
        \label{fig:sub1}
    \end{minipage}
    \hfill
    \begin{minipage}[t]{0.24\textwidth}
        \centering
        \includegraphics[width=\textwidth]{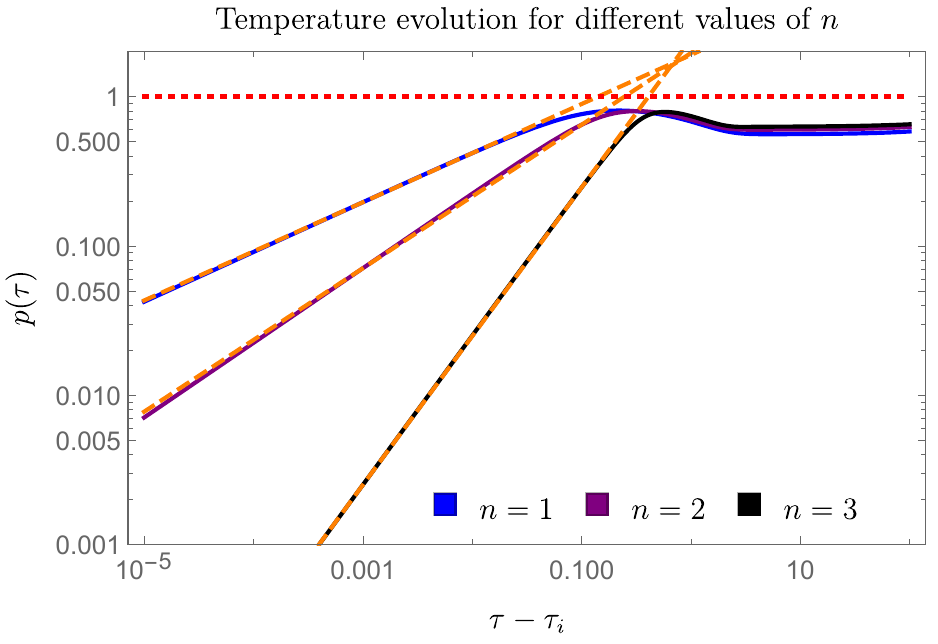}
        
        \label{fig:sub1}
    \end{minipage}
    \hfill   
    \begin{minipage}[t]{0.24\textwidth}
        \centering
        \includegraphics[width=\textwidth]{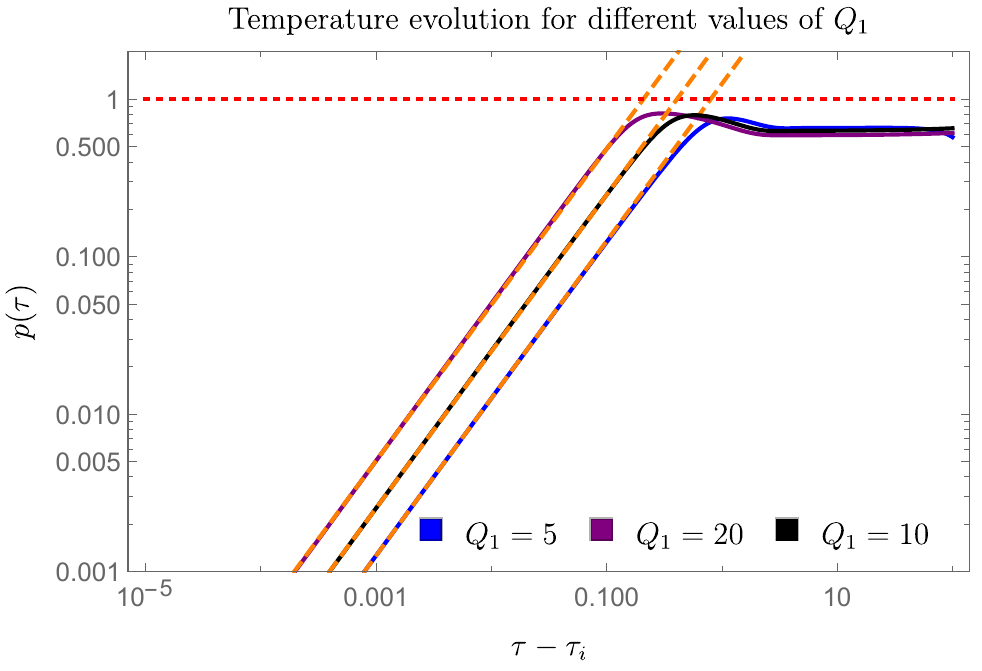}
        
        \label{fig:sub1}
    \end{minipage}
    \hfill    
     \begin{minipage}[t]{0.24\textwidth}
        \centering
        \includegraphics[width=\textwidth]{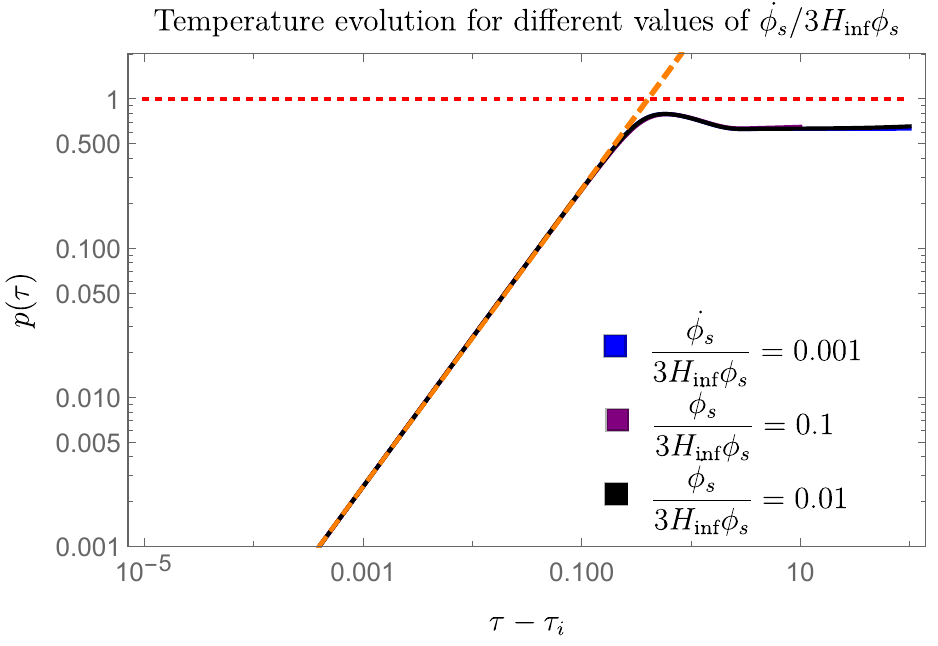}
        
        \label{fig:sub1}
    \end{minipage}
    \hfill
    \begin{minipage}[t]{0.24\textwidth}
        \centering
        \includegraphics[width=\textwidth]{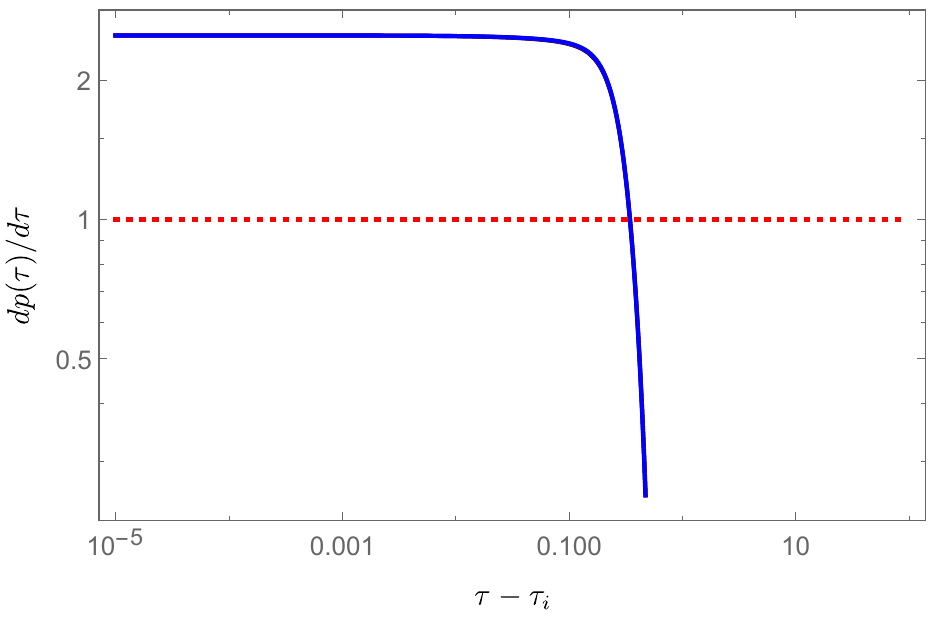}
       
        \label{fig:sub2}
    \end{minipage}
    \hfill
    \begin{minipage}[t]{0.24\textwidth}
        \centering
        \includegraphics[width=\textwidth]{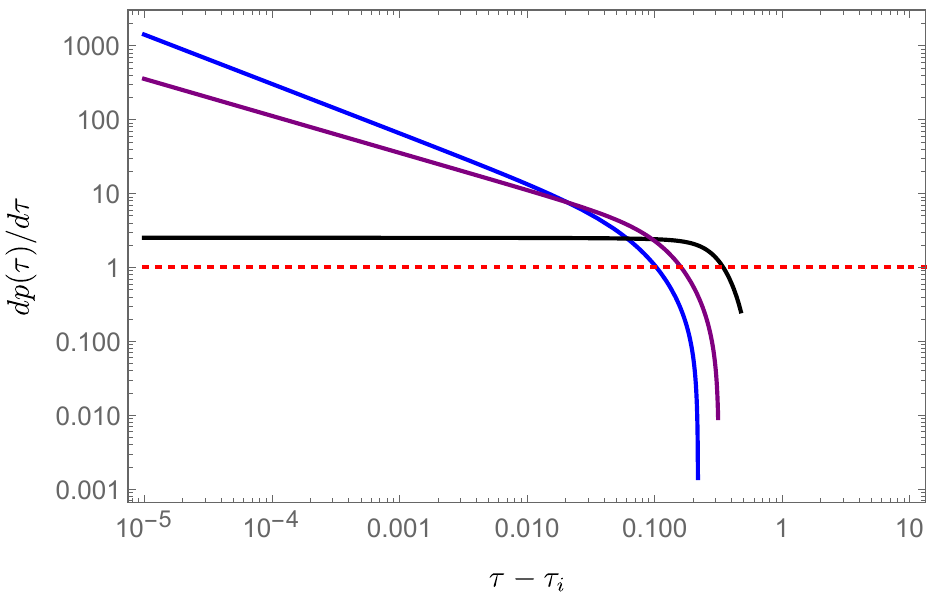}
        
        \label{fig:sub1}
    \end{minipage}
    \hfill
    \begin{minipage}[t]{0.24\textwidth}
        \centering
        \includegraphics[width=\textwidth]{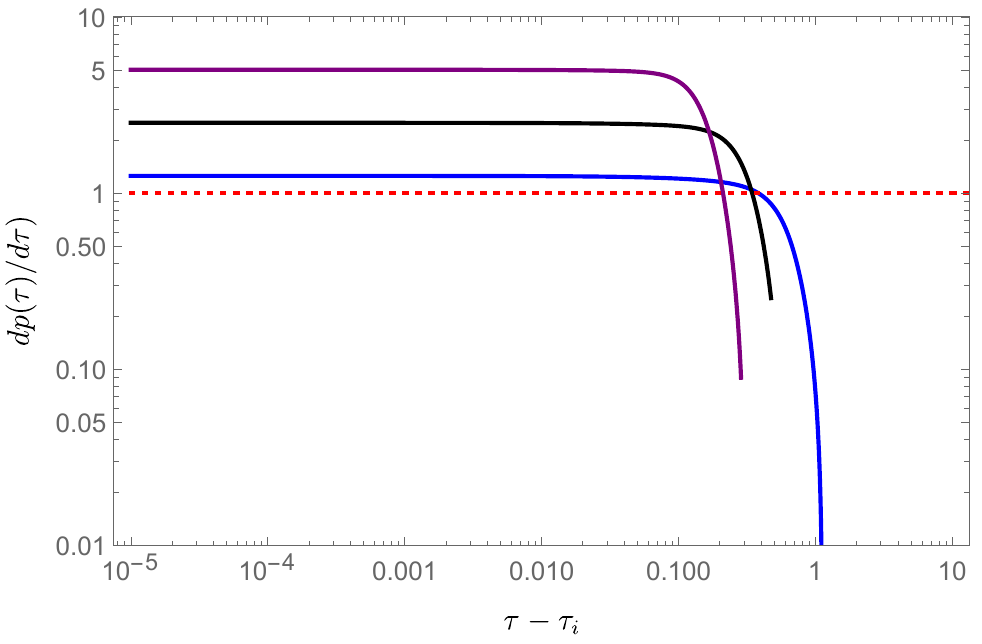}
        \label{fig:sub3}
    \end{minipage}
     \hfill
    \begin{minipage}[t]{0.24\textwidth}
        \centering
        \includegraphics[width=\textwidth]{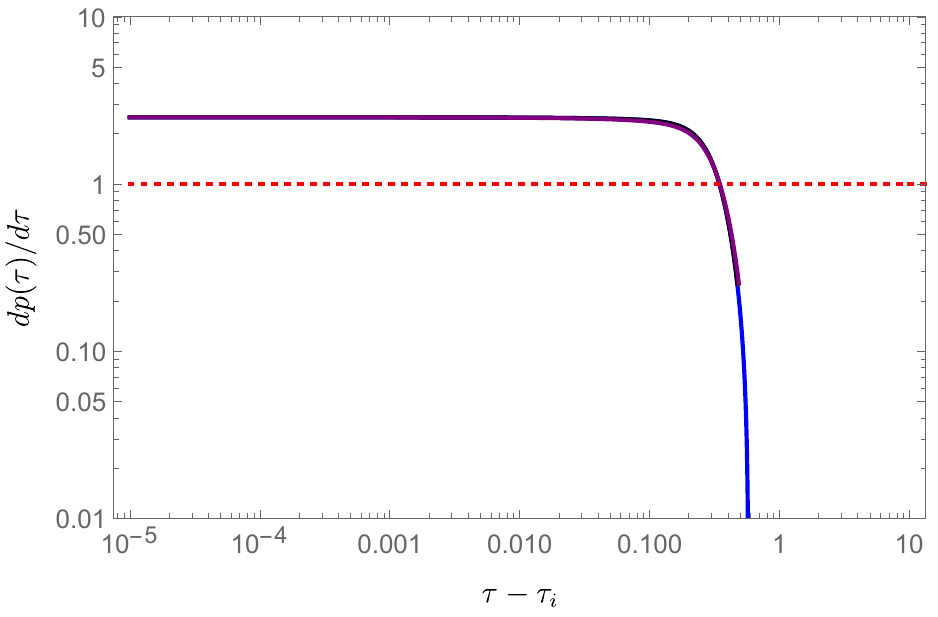}
        
        \label{fig:sub1}
    \end{minipage}
    \caption{This figure shows the evolution of \(p(\tau)\) (top panel) and $dp(\tau)/d\tau$ (bottom panel) under different parameter values during the heating phase. The black solid line corresponds to the parameter set: \(j=-2\), \(n=3\), \(Q_1=10\), and \(\frac{\dot{\phi}_s}{3H_{\text{inf}}\phi_s}=0.01\). The blue and purple solid lines respectively represent the evolution of \(p(\tau)\) (top panel) and \(p'(\tau)\) (bottom panel) with different parameters.}
    \label{figp}
\end{figure*}

In this subsection, we focus on the detailed temperature evolution dynamics during heating. 
Analysis of Eq.~\eqref{fun1} and \eqref{fun2} reveals that the thermal bath temperature variation primarily depend on the coupling coefficient \(Q\) and the kinetic energy of inflaton. Building on the framework of~\ref{subsec:IIA}, we assume that the Universe begins with a thermal bath with a low initial temperature $T_s \ll H$. 
The thermal friction gradually become significant and heat the system by transferring energy from the kinetic motion of inflaton to the thermal bath. 
The thermal bath temperature quickly stabilizes to an equilibrium value $T_{e}$, enabling a sustained WI phase. 

Initially, before heating begins, the thermal bath temperature \(T_s\) is very low which means the Universe is in the CI stage. During this stage, the inflaton satisfies: 
\begin{equation}
\label{evoi}
    3H_{\text{inf}}\dot{\phi} + V_{,\phi} = 0\,,
\end{equation} 
yielding an initial inflaton kinetic energy of \(|\dot{\phi}_s |=\frac{V_{,\phi}}{3H_{\text{inf}}}\) at the heating start. 
Upon the completion of the heating process, the Universe transitions to the WI phase, governed by:
\begin{equation}
\label{evoendphi}
    3H_{\text{inf}}(1 + Q)\dot{\phi}_e + V_{,\phi} = 0\,,
\end{equation}
\begin{equation}
\label{evoendT}
    4\overline{T_e}^4 = 3Q\dot{\phi}_{e}^2\,.
\end{equation}
Here, we apply several assumptions. 1.The term \(V_{,\phi}\) remains unchanged before and after heating. 2.The Hubble paramter \(H_{\text{inf}}\) is treated as a constant since the energy density of the thermal bath is negligible during WI.
3.The interaction is strong enough so that the thermal bath is in thermal equilibrium. Thus, the radiation energy density is given by \(\rho_r = \overline{T}^4\).

To facilitate solving Eq.~\eqref{fun1} \eqref{fun2}, we introduce dimensionless parameters \(x = \frac{\dot{\phi}}{\dot{\phi}_s}\), \(p = \frac{\overline{T}}{\sqrt{\dot{\phi}_s}}\), $y=\frac{\phi}{\phi_i}$, and \(\tau = 3H_{\text{inf}}t\), reducing Eq.~\eqref{fun1} and \eqref{fun2} to the compact forms:  
\begin{equation}
    \frac{dx}{d\tau} + (1 + Q)x + \frac{V_{,\phi}}{3H_{\text{inf}}\dot{\phi_s}} = 0\,,
\end{equation}
\begin{equation}
    \frac{dp}{d\tau} = \frac{1}{4} \frac{Qx^2}{p^3} - \frac{1}{3}p\,,
\end{equation}
\begin{equation}
    \frac{dy}{d\tau}=x\frac{\dot{\phi_s}}{3H_{\text{inf}}\phi_s}\,.
\end{equation}
As will be shown in the flowing, the heating process is extremely rapid, typically completing within a Hubble time. 
Consequently, during this stage, $V_{,\phi}$ can be approximated as a constant, which also supports the validity of our assumptions. 
Combining this assumption with Eq.~\eqref{evoi} yields $\frac{V_{,\phi}}{3H_{\text{inf}}\dot{\phi}_s} \approx -1$.
Moreover, $\frac{\dot{\phi}_s}{3H_{\text{inf}}\phi_s} \ll 1$ characterizes the relative rolling velocity of the inflaton during the CI, which is much smaller than unity.

The dissipation coefficient \(Q\) significantly influences the properties of WI, so we need to fix the form of $Q$. 
Various forms of $Q$ has been derived based on the first principles of quantum field theory. 
For example, in the two-stage mechanism model~\cite{Berera:1998gx,Yokoyama:1998ju}, \(Q \propto \phi^{-2}T^{2}\), which indicates that as the inflaton rolls, \(Q\) gradually increases. 
When dissipative effects become significant, the $\dot{\phi}$ largely decreases compared to the CI period, indicating that \(Q\) changes very slowly during WI. 
Another example is the distributed mass model~\cite{Berera:1999wt,Bastero-Gil:2018yen,Hall:2004zr}, where the dimensionless dissipation coefficient depends on a series of mass distributions of fields that coupled to the inflaton in the thermal bath.
Warm Inflation can also be naturally realized in the axion inflation scenario, a model in which the dissipation strength scales as \(Q \propto T^3\).~\cite{Berghaus:2019whh,Berghaus:2025dqi}
For generality, we adopt a power-law parameterization for the dissipation coefficient \(Q\), expressing it as a product of powers of the field \(\phi\) and the temperature \(T\)
\cite{Berera:2023liv,Kamali:2023lzq,Wang:2025duy}
\begin{equation}
    Q(\tau, p)=Q_1y^jp^n\,,
\end{equation}
where \(Q_1\), \(j\) and \(n\) are a model-dependent constants.

Figure~\ref{figp} shows the dependence of temperature evolution on the key parameters in the model. In each plot, as the thermal bath heats up, the temperature rises at an increasing rate. However, as it approaches its asymptotic maximum, the rise rate slows down rapidly and eventually stabilizes. Notably, in most stages, the temperature evolution approximately follows a power-law behavior.  
Furthermore, we find that the two parameters exerting the most significant influence on the heating process are \(n\) and \(Q_1\). First, a larger \(Q_1\) corresponds to a faster temperature increase rate: this is because a larger \(Q_1\) implies higher efficiency of energy transfer from the inflaton to the thermal bath. On the other hand, by combining Eq.~\eqref{evoi}, \eqref{evoendphi} and \eqref{evoendT}, we can approximately obtain \(T_e^4 \propto \frac{Q}{(1+Q)^2}\), which means the thermal bath temperature reaches its maximum when \(Q_1 \approx 1\) with the fact that $p\approx1$ and $y\approx1$ after the heating process. 
Thus, we observe that a larger \(Q_1\) corresponds to a lower \(T_e\) when \(Q_1 > 1\).  
The parameter \(n\) also strongly affects the temperature rise rate. For most WI models, \(0 < n \leq 3\). We find that a smaller \(n\) leads to a faster temperature rise rate. Particularly, when \(n = 3\), the temperature change rate remains constant in most stages of the heating process. 
In contrast to \(n\) and \(Q_1\), parameters \(j\) and $\frac{\dot{\phi_s}}{3H_{\text{inf}}\phi_s}$ play a negligible role in influencing temperature changes.

\section{Gravitational Wave signatures}
\label{sec3}
\subsection{GWs from first-order PTs}
In this subsection, we present the predictions of GWs from the first-order PTs both during inflation~(hPT) and after inflation~(cPT). 

In the early Universe, the existence of first-order PTs is commonly predicted by many physics models~\cite{Starobinsky:1982ee,Gross:1980br,Safronova:2017xyt}.
We assume a scalar field $\chi$ in the thermal bath can undergo such transitions, which is realized by the following effective potential at a finite temperature $T$~\cite{Quiros:1999jp,Linde:1981zj,Enqvist:1991xw}
\begin{figure}
\centering
\includegraphics[width=\linewidth]{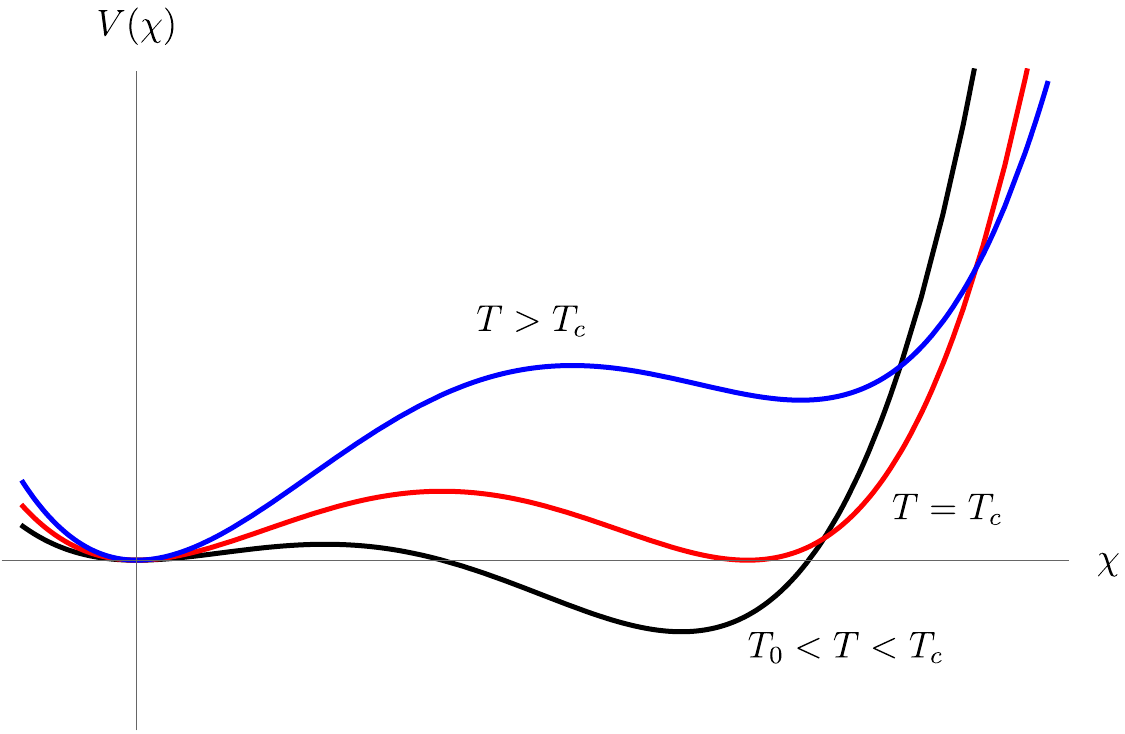}
\caption{The finite-temperature potential $V(\chi)$ for first-order PTs. The black, red, and blue curves correspond to the cases $T_0 < T < T_c$, $T = T_c$, and $T > T_c$, respectively.}
\label{Vchi}
\end{figure}

\begin{equation}
V(\chi,T) = \frac{\mu^2}{2} \left(T^2 - T_0^2 \right) \chi^2 - \frac{A}{3} T \chi^3 + \frac{\lambda}{4!} \chi^4,
\end{equation}
where $A$, $\mu$, $\lambda$, and $T_0$ are model-dependent parameters. The cubic term in the effective potential generated by thermal corrections is essential for realizing first-order PTs.
Figure~\ref{Vchi} illustrates $V(\chi)$ in the cases $T_0 < T < T_c$, $T = T_c$, and $T > T_c$.
The potential exhibits two minima, corresponding to the symmetric and broken phases. 
The two phases are degenerate at the critical temperature $T_c$. Above $T_c$, the symmetric phase is energetically favored, and $vice$ $versa$.

The thermal history of the Universe proceeds as follows.
Initially, the Universe is in a CI phase, with the thermal bath temperature $T_s$, satisfying $T_s< H < T_c$, where the $\chi$ field resides in the broken phase. 
During inflation, dissipative effects continuously heat the thermal bath, raising its temperature to $T_e$ and satisfies $T_e>T_c > H$, thereby effecting the transition from CI to WI.
During this process, the $\chi$ field undergoes a first-order hPT: the broken phase becomes metastable, and thermal fluctuations nucleate bubbles of the true vacuum. 
These bubbles expand and eventually percolate, completing the first-order PT.
Following inflation, the Universe naturally enters the radiation-dominated era without requiring an additional reheating phase. 
As the Universe continues to expand and cool, the temperature of the thermal bath eventually falls below $T_c$, rendering the symmetric phase metastable and triggering a subsequent first-order cooling phase transition.
In this way, the scalar fields $\chi$ undergoes both heating and cooling first-order PTs, each generating distinct gravitational wave signals whose observable features exhibit intriguing correlations.


Generally, GWs produced during such transitions originate from three primary sources: bubble collisions, sound waves, and magnetohydrodynamic turbulence.
The first source, bubble collisions, arises from the energy stored in the expanding bubble walls. 
The latter two sources result from the response of plasma to bubble nucleation and expansion~\cite{Kosowsky:1992rz,Hindmarsh:2015qta,Huber:2008hg}.

In the case of the cPT, the interactions between the PT field and the thermal bath produce a friction that counteracts bubble expansion, resulting in a terminal velocity $v<1$. 
In contrast, during the hPT, the transition proceeds from the broken to the symmetric phase, where it is energetically favorable for thermal bath particles to enter the bubbles. 
As plasma is drawn into the bubbles, the transfer of energy to the bubble walls accelerates them, potentially driving the walls into a runaway regime. 
Consequently, the energy available for GW production from bubble collisions is significantly larger during the hPT. 
For the cPT, bubbles do not typically reach runaway expansion, and the GW contribution from sound waves tends to dominate.
Additionally, because the hPT occurs during inflation, the resulted  GWs are redshifted by cosmic expansion. 
Their power spectrum also exhibits oscillatory features as a function of wavenumber.

We first consider GWs generated by bubble collisions during first-order hPTs that occur during inflation.
GWs generated during inflation undergo redshift as a result of exponential expansion, and their power spectra exhibit oscillations with wavenumber, thereby distinguishing them from GWs produced by post-inflationary first-order PTs. The distinction between inflationary GWs and non-inflationary GWs can be characterized by the deformation function, $S(f)$.
Using the envelope approximation, the present-day power spectrum of GWs generated by bubble collisions during first-order hPTs in inflation can be written as~\cite{Kosowsky:1992vn,Huber:2008hg,Cutting:2018tjt,Lewicki:2020azd,Lewicki:2022pdb}:
\begin{equation}
\label{omegagw1}
    h^2\Omega_{\text{GW,hPT}}(f)=h^2\hat{\Omega}_{\mathrm{GW}}\left(f\exp\left(N_*\right)\right)\frac{H^2_{\text{inf}}}{\frac{\pi^2 g_*}{90} \frac{T_e^4}{M^2_{\text{pl}}}}S(f)\,,
\end{equation}
where $N_*$ denotes the e-folding number at which GWs are generated by bubble collisions during the hPT, $h^2\hat{\Omega}_{GW}\left(f\exp\left(N_*\right)\right)$ describes the GW energy spectrum of GWs generated by the same source that have not undergone inflation, $g_*$ is degrees of freedom for radiation during WI.
~$f\exp\left(N_*\right)$ describes the redshift of GW frequency caused by inflation. 
Note that here we assume after the end of warm inflation, the Universe directly enters the radiation dominated era, with no other special cosmic phases such as the early matter dominated era. 
The term \(H_{\text{inf}}^2/\left(\frac{\pi^2 g_*}{90} \frac{T_e^4}{M_{\text{pl}}^2}\right)\) in Eq.~\eqref{omegagw1} can be treated as an amplification of $\hat{\Omega}_{\mathrm{GW}}$.
In WI, a thermal bath is sustained throughout the inflationary phase, eliminating the need for a separate reheating process. 
While the energy density fraction of inflaton is dominant during most of WI, it gradually decreases toward the end. 
When the inflaton and radiation energy densities become equal, the Universe transitions to radiation domination. 
This transition causes the Hubble parameter to drop from \(H_{\text{inf}}\) to \(\sqrt{\frac{\pi^2 g_*}{90}} \frac{T_e^2}{M_{\text{pl}}}\), reducing the critical density \(\rho_c \propto H^2\). 
A key assumption is that \(H_{\text{inf}} \gg \sqrt{\frac{\pi^2 g_*}{90}} \frac{T_e^2}{M_{\text{pl}}}\), ensuring radiation remains subdominant until the end of WI.

We define a new parameter $l\equiv\frac{2\pi a_0}{a_*H_{\text{inf}}}$, where $a_* \equiv a(N_*)$ and $a_0$ is the current scale factor, 
then the forms of the deformation function $S(f)$ and $h^2\hat{\Omega}_{GW}\left(f\exp\left(N_*\right)\right)$ are given as follows:
\begin{widetext}
    \begin{equation}
    \label{omegagw2}
    \begin{split}
        S(f)&=\left(\frac{\cos\left(lf(1-e^{-N_*})\right)}{l^2f^2}-\frac{\sin\left(lf(1-e^{-N_*})\right)}{l^3f^3}\right)^2\\
        &+lfe^{-N_*}\left(\left(\frac{1}{l^2f^2}+2e^{-N_*}-1\right)\frac{\sin\left(2lf(1-e^{-N_*})\right)}{l^4f^4}-\left(\frac{2-e^{-N_*}}{l^2f^2}+e^{-N_*}\right)\frac{\cos\left(2lf(1-e^{-N_*})\right)}{l^3f^3}+\frac{e^{-3N_*}}{lf}\left(\frac{1}{l^2f^2}+1\right)\right)\,,
    \end{split}
    \end{equation}

    \begin{equation}
    \label{omegagw3}
     h^2\hat{\Omega}_{\text{GW}}(f)=1.65\times10^{-5}\left(\frac{100}{g_*}\right)^{1/3}\left(\frac{H_{\text{inf}}}{\beta_{\text{hPT}}}\right)^2\left(\frac{\kappa_{\text{hPT}}\alpha_{\text{hPT}}}{1+\alpha_{\text{hPT}}+\mathcal{R}_{\text{hPT}}}\right)^2\left(\frac{0.11v_{\text{hPT}}^3}{0.42+v_{\text{hPT}}}^2\right)\frac{3.8\left(f/f_{\text{hPT}}\right)^{2.8}}{1+2.8\left(f/f_{\text{hPT}}\right)^{3.8}}\,.
    \end{equation}
\end{widetext}

The inverse of characteristic time scale of hPTs, \(\beta_{\text{hPT}}\), can generally be expressed as:  
\begin{equation}
    \beta_{\text{hPT}}\approx\hat{\beta}_{\text{hPT}} \left|\frac{d\ln T}{dt}\right|\,,
\end{equation}
where $\hat{\beta}_{\text{hPT}}$ is a model-dependent parameter determined by the Euclidean action of PTs, while the temperature change rate $|d\ln T/dt|$ reflects the heating dynamics which we have discussed in detail in the Section~\ref{sec2}.
For the PT to complete efficiently within inflation, its timescale must be shorter than the Hubble time, implying $\beta_{\text{hPT}} / H_{\text{inf}} > 1$, and this condition is readily satisfied in our scenario.

The parameter $\alpha_{\text{hPT}}$ represents the ratio of the vacuum energy density of the $\chi$ field to the energy density of the thermal bath, and typically $\alpha_{\text{hPT}} \ll 1$. 
The ratio $\mathcal{R}_{\text{hPT}} = \rho_\phi / \rho_r$ quantifies energy density ratio relative to the inflaton and the radiation. Since inflation is dominated by the inflaton, $\mathcal{R}_{\text{hPT}} \gg 1$. 
The efficiency factor $\kappa_{\text{hPT}}$ characterizes the fraction of the vacuum energy that is converted into bubble wall energy relevant for GW production. 
In hPTs, energy is transferred into the bubble walls via the anti-friction of the thermal bath~\cite{Buen-Abad:2023hex} and allowing $\kappa_{\text{hPT}}$ to potentially exceed unity, unlike in cPTs where $\kappa \leq 1$. Due to the runaway expansion of the bubbles, the bubble wall velocity $v_{\text{hPT}}$ approaches unity.
Finally, \(f_{\text{hPT}}\) is the peak frequency of the spectrum contribution from bubble collisions, which is redshifted due to cosmic expansion, corresponding to the present-day value
\begin{widetext}
    \begin{equation}
    \label{omegagw4}
    f_{\text{hPT}}=16.5\times10^{-3}\mathrm{mHz}\left(\frac{0.62}{1.8-0.1v_{\text{hPT}}+v_{\text{hPT}}^2}\right)\left(\frac{\beta_{\text{hPT}}}{H_{\text{inf}}}\right)\left(\frac{T_e}{100\mathrm{GeV}}\right)\left(\frac{g_*}{100}\right)^{1/6}\,,
\end{equation}
\end{widetext}

For the cPT, the dominant contribution to GWs comes from sound waves, and the present-day energy spectrum of GWs is given by
\begin{widetext}
    \begin{equation}
    \label{omegagw5}
        h^2\Omega(f)_{\text{GW,cPT}}=2.56\times10^{-6}\left(\frac{H_*}{\beta_{\text{cPT}}}\right)\left(\frac{\kappa_{\text{cPT}}\alpha_{\text{cPT}}}{1+\alpha_{\text{cPT}}}\right)^2\left(\frac{100}{g_*}\right)^{1/3}v_{\text{cPT}}\left(\frac{f}{f_{\text{cPT}}}\right)^3\left(\frac{7}{4+3\left(f/f_{\text{cPT}}\right)^{2}}\right)^{7/2}\,,
    \end{equation}
\end{widetext}
where \(H_* = \sqrt{\dfrac{1}{3M^2_{\text{pl}}} \dfrac{\pi^2}{30} g_* T_c^4}\) corresponds to the Hubble parameter at the time of GW generation under the assumption that the cosmic energy density is dominated by radiation. 
Since the cPT also typically proceeds rapidly, for simplicity, we approximate the temperature of the thermal bath at the time of GW generation as $T_c$.
 \(\beta_{\text{cPT}}\) is the inverse of the characteristic time scale for cPTs. We assume that cPT occurs in the radiation dominated era after the end of WI, therefore, $\beta_{\text{cPT}}$ can be expressed as
\begin{equation}
    \beta_{\text{cPT}}\approx\hat{\beta}_{\text{cPT}} \left|\frac{d\ln T}{dt}\right|\approx\hat{\beta}_{\text{cPT}}H_*\,.
\end{equation}
Here, \(\hat{\beta}_{\text{cPT}}\) is a model-dependent parameter close to \(\hat{\beta}_{\text{hPT}}\),
and \(\left|\frac{d\ln T}{dt}\right|\) describes the cooling rate of the thermal bath, $H_*$ is the Hubble parameter during the cPT. 
\(\alpha_{\text{cPT}}\) denotes the ratio of the vacuum energy of the PT field \(\chi\) to the radiation energy density during the cPT. 
\(\kappa_{\text{cPT}}\) characterizes the fraction of energy released during the cPT that is converted into GWs. 
Using the bag model~\cite{Lewicki:2022pdb,Weir:2016tov} 
and corresponding fits for \(\kappa_{\text{cPT}}\), 
we have \(\kappa_{\text{cPT}} \approx 5\alpha_{\text{cPT}} v_{\text{cPT}}^{6/5}\), where \(v_{\text{cPT}}\) is the expansion velocity of bubble walls. Frictional forces from the thermal bath balance the expansion force of the new phase, causing the bubble walls to move at a constant~\cite{Espinosa:2010hh}. 
\(f_{\text{cPT}}\) is the peak frequency of the spectrum contribution from sound waves, corresponding to the present-day value

\begin{equation}
\label{omegagw6}
    f_{\text{cPT}}=1.9\times10^{-2}\mathrm{mHz}\frac{1}{v_{\text{cPT}}}\left(\frac{\beta_{\text{cPT}}}{H_*}\right)\left(\frac{T_c}{100\mathrm{GeV}}\right)\left(\frac{g_*}{100}\right)^{1/6}\,.
\end{equation}

\subsection{Detectability of GWs and Physical Implications}

\begin{figure}
    \centering
    \includegraphics[width=1\linewidth]{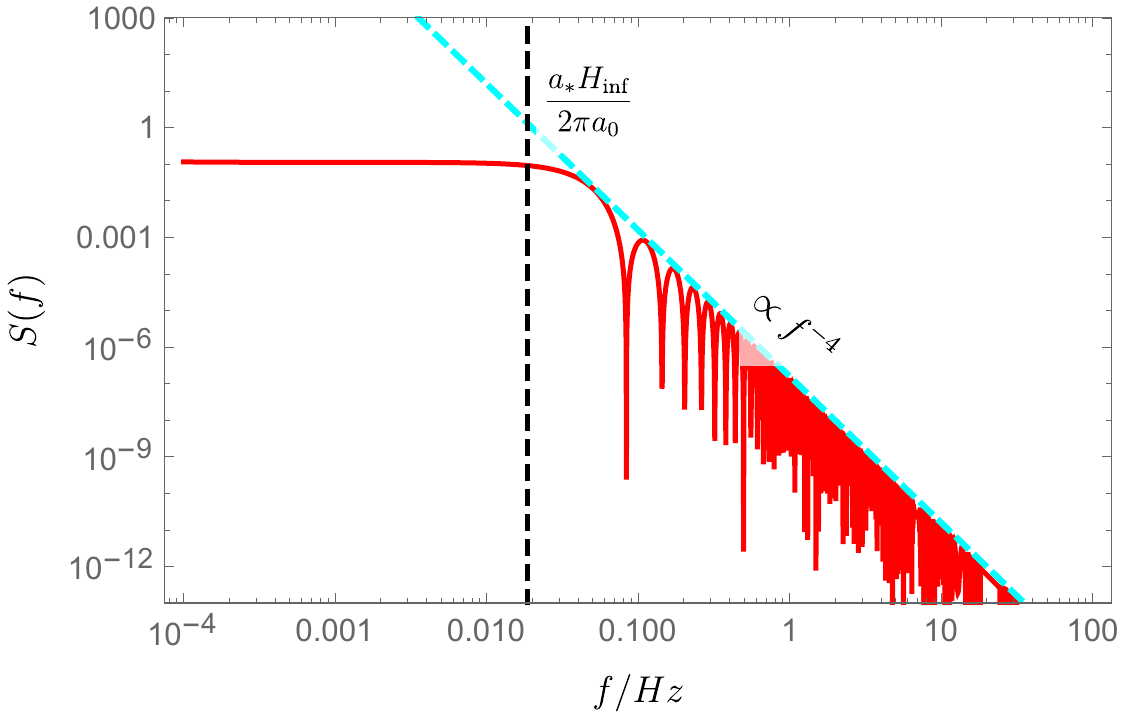}
    \caption{The deformation equation is given for the parameter set \( N_* = 18 \), \( T_e =10^{14} \, \text{GeV} \), $T_c=10^{13}\text{GeV}$ and \( H_{\text{inf}} = 10^{12} \, \text{GeV} \). 
    Here, the black dashed line represents the frequency corresponding to the Hubble scale during hPT.}
    \label{figs}
\end{figure}
\begin{figure}
    \centering
    \includegraphics[width=1\linewidth]{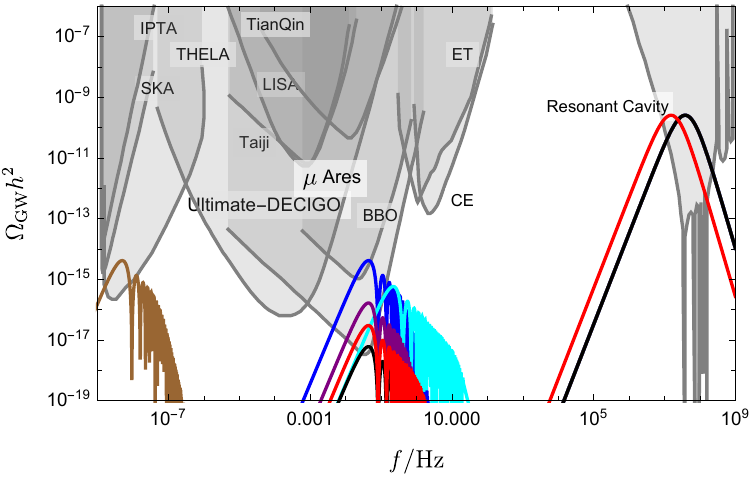}
    \caption{The energy spectrum of GWs generated by first-order PTs in the WI scenario when we select different parameter values.
    the GWs with an oscillatory structure at lower frequencies originate from hPT, while those at higher frequencies originate from cPT.
    The blue solid line represents the parameter set where $T_e=10^{14}\mathrm{GeV}$, $T_c=5\times10^{13}\mathrm{GeV}$, $H_{\text{inf}}=10^{10}\mathrm{GeV}$, $N_*=18$, $Q=5$, and the purple, red, cyan, black, and brown solid lines represent the parameter sets where \(H_{\text{inf}}=5\times10^{10}\mathrm{GeV}\), \(T_{c}=10^{13}\mathrm{GeV}\), \(T_{e}=5\times10^{14}\mathrm{GeV}\), \(Q=10\), and \(N_*=34\) respectively, with all other parameters consistent with those of the blue solid line.  }
    \label{figgwtot}
\end{figure}
\begin{figure}
    \centering
    \includegraphics[width=1\linewidth]{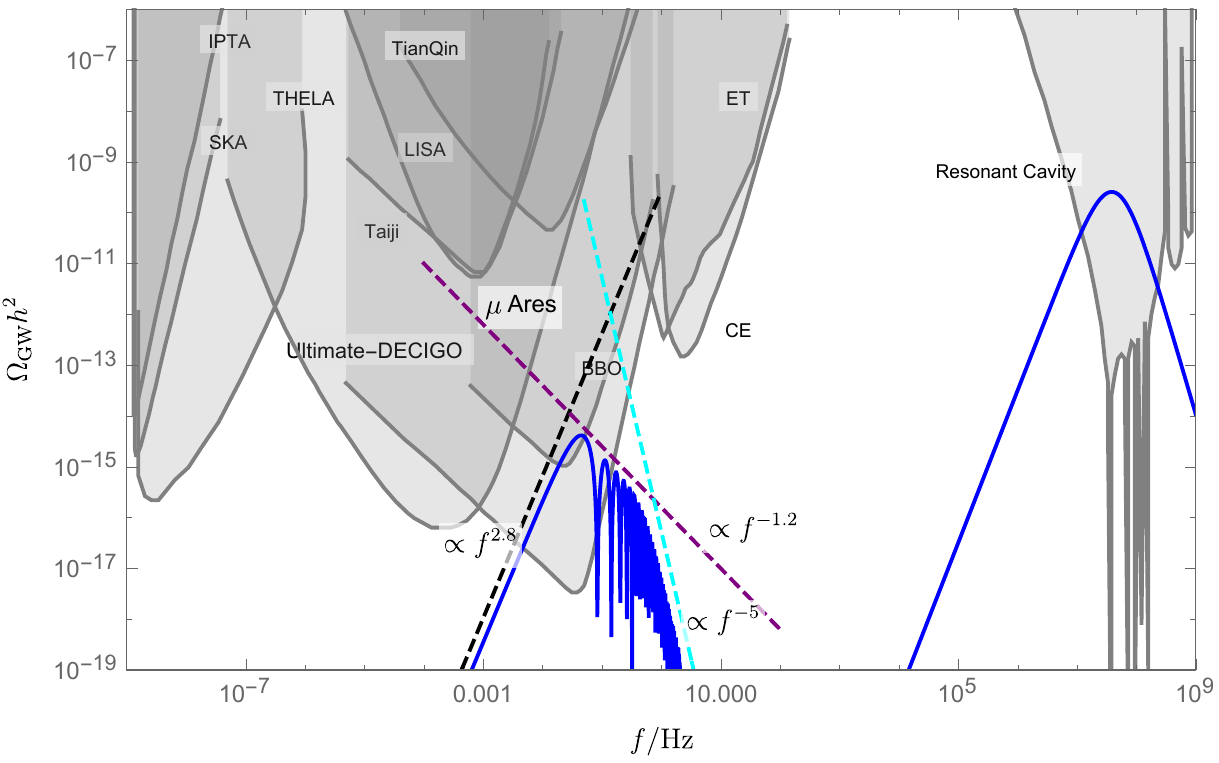}
    \caption{The energy spectrum of GWs generated by first-order PTs in the WI scenario when we select $T_e=10^{14}\mathrm{GeV}$, $T_c=5\times10^{13}\mathrm{GeV}$, $H_{\text{inf}}=10^{10}\mathrm{GeV}$, $N_*=18$.
    }
    \label{figgw}
\end{figure}

This section primarily discusses the detectability of GWs induced by hPTs and cPTs in the WI scenario. 
To show the result, we adopt minimal warm inflation~\cite{Berghaus:2019whh}. 
In this scenario, the inflaton is an axion-like field, i.e., \(j = 0\) and \(n = 3\), as we discussed in Sec.~\ref{sec:IIB}.  
Note that the mechanism of this work is also applicable to other inflationary models.

Fig.~\ref{figgw} illustrates the dependence of GWs generated from FOPTs on the parameters discussed in this paper, and we find that the rate of temperature change exerts a crucial influence on GWs generated by hPTs, an effect that is primarily governed by the value of $\beta_{\text{hPT}}$. 
First, as shown in Eq.~\eqref{omegagw3}, the value of $\beta_{\text{hPT}}$ directly governs the power spectrum of GWs at the time of their generation during hPTs. 
Second, as showed in Eq.~\eqref{omegagw4}, the value of $\beta_{\text{hPT}}$ also dictates the scale at which GWs are generated, i.e., a larger $\beta_{\text{hPT}}$ corresponds to a smaller scale of GWs. 
Since hPTs occur during the WI, GWs generated by hPTs on subhorizon scales will exit the horizon, become frozen as superhorizon perturbations, and later re-enter the horizon during the subsequent radiation-dominated era.
Before exiting the horizon, small-scale GWs generated by hPTs experience more pronounced attenuation by expansion, where the GW energy spectrum is proportional to $f^{-4}$, as illustrated in Fig.~\ref{figs}. Through the combination of Eq.~\eqref{omegagw3} and \eqref{omegagw4}, the dependence of the power spectrum of GWs on the rate of temperature change can be approximately derived as $\Omega_{\text{GW}} \propto \left(\frac{\ln T}{dt}\right)^{-6}$. 


The power spectrum of GW from cPT is primarily determined by \(\beta_{\text{cPT}}\), 
while the frequency band of $\Omega_{\text{GW,cPT}}$ is governed by the critical temperature \(T_c\). 
Since first-order cPTs are predicted in a wide range of early-Universe models, the additional GW signals from hPTs in lower frequency bands
serves as a benchmark signal to differentiate our model from others.
Note that in the CI scenario, hPTs could also occur during the reheating phase. In this case, both the hPT and the cPT take place after inflation, with the hPT occurring earlier. Since the hPT corresponds to a larger critical density (\(\rho_{c}\)), it produces GWs at a higher frequency than those from the cPT.
In contrast, hPTs in the WI scenario occur during the inflationary epoch, the generated GWs undergo redshift due to cosmic expansion, ultimately resulting in their power spectra lying in much lower frequency bands. 
Additionally, the power spectra of GW from hPTs in WI exhibit oscillatory structures, which is also a key distinguishing feature from CI scenarios.  

The frequency of GWs generated by hPTs, while influenced by parameters like \(T_e\), \(\beta_{\text{\text{hPT}}}\) and $H_{\text{inf}}$, is predominantly determined by the number of e-folds \(N_*\), due to an exponential dependence. 
This is because the GW modes exit the horizon during inflation and are subsequently ``frozen" until they re-enter during the radiation-dominated era. 
Given the lack of strict observational constraints on \(N_*\), the resulting GW signals could potentially be detected across a wide frequency range by observers such as BBO, Ultimate-DECIGO, or SKA.

This GW signal generated by first-order PT with double-peak structure can serve as a probe of the early Universe, helping us understand the thermal history of the early Universe. 
In the scenario we discuss, the quantities of interest include inflationary Hubble parameter $H_{\text{inf}}$, the thermal bath temperature $T_e$ during WI, the critical temperature $T_c$ of PT, the e-folding number $N_*$ at which the heating PT occurs, and the temperature change rate during WI, $\frac{d\ln T}{dt}$—or equivalently, $\beta_{\text{hPT}}$.  
We can roughly determine the critical temperature $T_c$ using  GWs generated by cPT, while the other parameters can be constrained by comparing the two peaks of the GW signal.  
First, by examining Fig.~\ref{figgw}, we find that the GW signal from PT during WI not only exhibits an oscillatory feature but also shows distinct frequency-dependent trends: for $f < \frac{a_* H_{\text{inf}}}{2\pi a_0}$, $\Omega_{\text{GW,hPT}} \propto f^{2.8}$; for $\frac{a_* H_{\text{inf}}}{2\pi a_0} < f < f_{\text{hPT}} \exp(-N_*)$, $\Omega_{\text{GW,hPT}} \propto f^{-1.2}$; and for $f > f_{\text{hPT}} \exp(-N_*)$, $\Omega_{\text{GW,hPT}} \propto f^{-5}$.  
Here, $\frac{a_* H_{\text{inf}}}{2\pi a_0}$ and $f_{\text{hPT}} \exp(-N_*)$ describe the Hubble scale during WI and the characteristic scale of bubble collisions during hPT, which satisfy $\frac{a_* H_{\text{inf}}}{2\pi a_0}\propto \exp(-N_*) \frac{T_0}{T_e} H_{\text{inf}}$ and $f_{\text{hPT}} \propto\frac{\beta_{\text{hPT}}}{H_{\text{inf}}} T_e$, respectively, where $T_0$ is the current temperature of the thermal bath.  
On the other hand, the peak of the GW signal from the hPT depends on $\frac{H_{\text{inf}}}{\beta_{\text{hPT}}} \frac{\alpha^2}{\mathcal{R}^2} \frac{H_{\text{inf}}^2 M_{\text{pl}}^2}{T_e^4}$, where $\alpha$ can be constrained by GWs generated by cPT.  
Note that we use the fact that $\mathcal{R}\gg1\gg\alpha$ here.
Through observations of GWs, we can obtain constraints on \(T_e\), \(H_{\text{inf}}\), and \(\beta_{\text{hPT}}\), and these constraints exactly help us determine the value of \(\frac{\beta_{\text{hPT}}}{H_{\text{inf}}}\).  
As mentioned earlier, $\frac{\beta_{\text{hPT}}}{H_{\text{inf}}}$ depends primarily on the temperature change rate of the thermal bath during the heating process of WI. 
Thus, through observations of GWs generated by first-order PT, we can not only distinguish whether inflation is WI but also gain deeper insights into the generation of the thermal bath. 
On the other hand, the relationship between the frequencies corresponding to the two peaks of the GW signal mainly depends on $N_*$ and $T_e$. Due to the presence of $\exp(N_*)$, the influence of $N_*$ is dominant. 
Thus, comparing the two peaks can help us determine the duration of WI.

\section{Conclusions} \label{conclusion}
\label{sec5}
This paper investigates a novel mechanism within the WI framework, where the thermal bath is heated during inflation, triggering a hPT. After inflation, as the cosmic temperature gradually decreases, the same model also gives rise to a cPT. We predict GW signals from both PTs, with the hPT-generated GWs having a much lower peak frequency due to redshifting from the exponential expansion during inflation. Upcoming GW observatories such as DECIGO, BBO, and pulsar timing arrays may detect these signals. Multiband GW observations could allow simultaneous detection of GWs from both hPTs and cPTs. The correlation between their peak frequencies and amplitudes might provide valuable insights into dissipative dynamics and the thermal history of the early Universe, offering a promising way to distinguish WI from CI and to probe the WI parameter space more precisely.

The frequency of the GW signal is primarily determined by two factors, $N_*$ and $T_c$. 
In particular, $T_c$ directly sets the frequency of GWs sourced by cPTs, whereas $N_*$ mainly controls the frequency difference between GWs originating from cPTs and hPT. 
The amplitude of GWs generated by hPTs is closely linked to WI parameters and the early-Universe thermal history, especially the dissipation coefficient $Q$, the critical temperature $T_c$ of PTs, the thermal bath temperature $T_e$ during WI, and the inflationary scale $H_{\text{inf}}$.

This mechanism is especially effective in the weak dissipation regime. In contrast, when dissipation is strong ($Q \gg 1$), the first-order hPT occurs very rapidly, producing weak GW signals that are challenging to detect. However, it significantly slows the rolling speed of the inflaton, thereby enhancing scalar perturbations, which can be constrained through observations of curvature perturbations and scalar-induced GWs.
This opens a promising direction for future research, reinforcing the theoretical underpinnings of WI and improving its observational testability.

It should be noted that when calculating the current GW power spectrum, we approximately assume the Universe undergoes adiabatic expansion after the end of inflation in this paper. 
If there is a radiation reheating process, such as Q-ball or PBH decay, in the early stage of the radiation-dominated era before the temperature drops to \(T_c\), and that increases the radiation temperature by a factor of \(\gamma\), then GWs induced by hPT will be suppressed by \(\gamma^{-4}\). 
Since this reheating process does not affect the redshift of GWs induced by cPT, the gap between the two GW peaks will widen further. 
Additionally, while this process causes redshift of the GW spectrum, its influence is secondary compared to that of \(N_*\).


This work assumes that bubble collisions are the dominant source of GWs generated by hPTs, with their spectrum described by the envelope approximation. 
Although this assumption is well-motivated, it calls for validation through dedicated numerical simulations. 
Furthermore, the analysis presumes the existence of a thermal bath in near-equilibrium throughout the heating process; in scenarios of weak dissipation, however, non-equilibrium effects must be taken into account. 
Finally, although this study does not systematically examine different inflationary potential functions, it should be emphasized that certain inflationary potential functions with special structures, such as those enabling ultra-slow-roll inflation, may give rise to distinctive GW signatures.
These open issues will be addressed in future work.

\begin{acknowledgments}
    This work is supported in part by the National Key Research and Development Program of China Grants No. 2020YFC2201501 and No. 2021YFC2203002, in part by the National Natural Science Foundation of China Grants No. 12588101, No. 12147103, No. 12235019, No. 12075297 and No. 12147103, in part by the Science Research Grants from the China Manned Space Project with No. CMS-CSST-2021-B01, in part by the Fundamental Research Funds for the Central Universities.
\end{acknowledgments}

\bibliography{citeLib}
\end{document}